\documentclass{IOS-Book-Article}

\usepackage{mathptmx}
\usepackage{soul}\setuldepth{article}
\def\hb{\hbox to 11.5 cm{}}
\usepackage{amsmath}
\usepackage{scrextend}
\usepackage{graphicx}
\usepackage{comment}
\usepackage{caption}
\usepackage{subcaption}
\usepackage{hyperref}
\hypersetup{
    colorlinks=true,
    linkcolor=blue,
    filecolor=magenta,      
    urlcolor=cyan,
    pdftitle={Overleaf Example},
    pdfpagemode=FullScreen,
    }
\usepackage{wrapfig}

\begin{document}

\pagestyle{headings}
\def\thepage{}
\begin{frontmatter}              

%\pretitle{Pretitle}
\title{Human-AI Collaboration for Estimating Scientific Replicability}

\author[A]{\fnms{Tatiana} \snm{Chakravorti}}, 
\author[A]{\fnms{Robert} \snm{Fraleigh}},
\author[A]{\fnms{Timothy} \snm{Fritton}},
\author[A]{\fnms{Christopher} \snm{Griffin}},
\author[A]{\fnms{Vaibhav} \snm{Singh}},
\author[A]{\fnms{Sai} \snm{Koneru}},
\author[A]{\fnms{C. Lee} \snm{Giles}},
\author[B]{\fnms{David} \snm{Pennock}},
\author[A]{\fnms{Anthony} \snm{Kwasnica}},
\author[A]{\fnms{Sarah} \snm{Rajtmajer}\thanks{Corresponding Author: smr48@psu.edu}}

%\runningauthor{T. Chakravorti et al.}

\address[A]{The Pennsylvania State University}
\address[B]{Rutgers University}

\begin{abstract}
Determining whether published scientific findings can successfully be replicated is a long-standing challenge in the empirical sciences. Existing approaches for replicability assessment typically rely either on human judgment, i.e., creative assembly of human experts, or on machine learning models trained on paper content metadata. While both approaches have demonstrated value, each also has important limitations. Human forecasts can be influenced by cognitive biases and narrow exposure to the research literature, while automated assessments often struggle to capture contextual cues and subtle signals of credibility. In this paper, we examine a hybrid approach. Specifically, we introduce a hybrid prediction market in which algorithmic agents trade alongside human participants to jointly estimate the likelihood that a published scientific finding will be corroborated via the outcome of a controlled replication study. Agents are trained on outcomes from hundreds of prior replication studies while human participants contribute domain knowledge through real-time trading. We evaluate this hybrid approach through multiple live experiments involving participants from different academic disciplines and compare its performance to artificial-only and human-only baselines. Our results show that, except for a few cases, hybrid markets match or outperform artificial prediction markets, producing more accurate and reliable replication forecasts.
\end{abstract}

\end{frontmatter}

\section{Introduction}
In recent decades, artificial intelligence (AI) and machine learning (ML) have achieved substantial advances, reaching or surpassing human level performance on a wide range of complex tasks. However, despite these successes, the deployment of AI systems in real-world decision-making remains constrained by important limitations. While today's models perform well on standardized benchmarks, such performance may reflect statistical pattern matching rather than robust, generalizable reasoning \cite{chizhov2025hellaswag, spiesberger2026soft, chen2025benchmarking}. As a result, they often struggle with tasks requiring counterfactual reasoning, contextual understanding, and learning from limited data \cite{kim2025limitations, dellermann2021future, puig2020watch}, and may produce confident but incorrect outputs under distribution shift \cite{alansari2025large, lin2026omitted}. These limitations restrict the use of AI in high-stakes settings where reliability and interpretability are critical.

At the same time, human judgment provides complementary strengths, including contextual reasoning, domain expertise, and the ability to interpret incomplete or ambiguous information. Rather than viewing human and artificial intelligence as substitutes, a growing body of research emphasizes their potential to function as complementary components of hybrid systems that integrate human insight with computational scalability \cite{dellermann2019hybrid, wang2020human}. A central challenge of this work is how to effectively combine heterogeneous sources of information, e.g., human beliefs and algorithmic predictions, into coherent and well-calibrated forecasts.

Prediction markets offer a natural framework for this integration. By allowing participants to buy and sell contracts tied to future outcomes, prediction markets aggregate distributed information into probabilistic forecasts through market prices \cite{Arrow:2008, hanson2006information, manski2006interpreting}. Artificial prediction markets extend this idea to computational settings. In these systems, algorithmic agents participate in a simulated market and interact through market-based mechanisms to produce predictions  \cite{barbu2012introduction}. Prior work has demonstrated that artificial prediction markets can function as supervised learning algorithms, achieving performance comparable to standard classification methods on select tasks \cite{barbu2012introduction, barbu2013artificial, nakshatri2021design}. However, the effectiveness of artificial prediction markets is highly sensitive to design choices such as liquidity, market duration, and agent initialization. Moreover, like their human counterparts, artificial markets can suffer from insufficient participation or engagement \cite{Arrow:2008, tetlock2008liquidity, rajtmajer2022synthetic}. As a result, artificial prediction markets have not yet presented a robust alternative to conventional supervised learning techniques.

Rather than viewing these limitations as reasons to abandon market-based approaches, we argue that they point toward an opportunity for human-AI collaboration. In hybrid prediction markets, human participants trade alongside trained algorithmic agents, allowing human and AI information and judgment to be dynamically aggregated into probabilistic forecasts. This design leverages the speed and breadth of algorithmic agents while incorporating human contextual insight and expertise. We investigate whether hybrid markets can improve predictive performance on complex evaluative tasks.

We study this question in the domain of forecasting scientific replication outcomes, a task that has proven challenging for both human and automated methods. Prior research has explored human crowd-based approaches, including prediction markets and structured expert elicitation \cite{liu2020replication, fraser2023predicting}, as well as machine learning models trained on paper metadata and statistical features \cite{altmejd2019predicting, yang2020estimating, pawel2020probabilistic, wu2021predicting}. Despite these efforts, neither class of approaches has consistently achieved high predictive accuracy. Human forecasts are shaped by cognitive biases and limited familiarity with the broader literature, while automated models often fail to capture nuanced credibility signals embedded within scientific work. These complementary limitations make replication forecasting an ideal testbed for evaluating hybrid human-AI methodologies. 

Our work is scaffolded by the following research questions:

\noindent \textbf{RQ1:} Are hybrid human-AI prediction markets able to produce more accurate replication forecasts? How do these forecasts compare to human-only or agent-only markets?

\noindent \textbf{RQ2:} What information and strategies do human participants use when participating in a replication prediction market? %Does the AI participation increase? Do humans change their decisions?

Through a series of live experiments involving human participants from corresponding disciplinary backgrounds, we compare hybrid markets against artificial-only and human-only baselines in the context of replication prediction.  Our findings demonstrate that hybrid markets consistently match or outperform artificial markets, with notable exceptions (in marketing and education). While performance varied by domain, hybrid markets generally achieved lower or comparable mean absolute error and demonstrated more stable aggregation of information. Survey evidence indicates that human traders primarily relied on epistemic beliefs rather than purely profit-maximizing strategies, i.e., hybrid markets integrate both algorithmic pattern recognition and domain-informed human judgment. Our findings suggest that thoughtfully designed hybrid human-AI systems can improve reliability in complex scientific evaluation tasks, offering a scalable framework for strengthening evidence assessment in research ecosystems. 

\section{Related Work}

This research builds upon and contributes to several distinct literatures. The first is approaches to estimate or predict the replicability of scientific findings. The second is prior work on artificial prediction markets and market-based machine learning systems that formalize probability estimation as a trading process among algorithmic agents. The third is emerging literature on human-AI collaboration in forecasting and evaluative judgment. 

\subsection{Replication prediction}
Large-scale replication\footnote{A note on terminology: By \emph{reproducibility}, we refer to computational repeatability – obtaining consistent computational results using the same data, methods, code, and conditions of analysis; by \emph{replicability}, mean obtaining consistent results on a new dataset using similar methods. We adopt these definitions from \cite{national2019reproducibility}.} projects in psychology \cite{nosek349corresponding}, economics \cite{camerer2016evaluating}, sociology \cite{camerer2018evaluating}, biology \cite{errington2014open}, physics \cite{feger2019designing} and beyond have turned up disappointing results, putting questions of scientific credibility at the forefront of scientific debate. Because direct replications are costly and slow, replication prediction methods aim to prioritize which studies to replicate, allocate verification resources, and provide decision-makers with calibrated confidence signals about research claims. Recent work has converged on three complementary approaches: 1) human forecasting; 2) algorithmic prediction using reported statistics and paper metadata; and 3) text-based machine learning that uses features of the writing itself, sometimes combined with other signals. A robust line of research shows that researchers’ aggregate beliefs contain meaningful information about scientific replicability. This has been operationalized through forecasting surveys and prediction markets, where participants estimate the likelihood that a study will successfully replicate \cite{chandrashekar2026using}. Prior research has established crowd-based forecasting as one of the two dominant paradigms for predicting replication outcomes, including structured elicitation methods and prediction markets \cite{liu2020replication, fraser2023predicting}. Collective forecasting principles extend beyond human participants. Recent work demonstrates that aggregated large language model (LLM) “crowds” can match or approach the accuracy of human forecasting crowds \cite{schoenegger2024wisdom}. However, large-scale analyses of crowd forecasts also reveal systematic calibration challenges and information gaps. These limitations constrain predictive performance and reflect human cognitive biases as well as restricted domain knowledge \cite{dellavigna2025forecasting}. Alongside crowd-based forecasting, a second dominant paradigm relies on machine learning models trained on structured features extracted from published studies. These models use statistical properties, methodological indicators, and other coded characteristics to estimate replication likelihood \cite{altmejd2019predicting, yang2020estimating, pawel2020probabilistic, wu2021predicting}. Machine learning approaches aim to scale replication prediction beyond what is feasible with human forecasting alone, offering automated and potentially generalizable prediction systems across large corpora of scientific articles. Extending machine learning approaches further, research has developed large-scale text-based models to estimate replication likelihood directly from article text. A prominent large-scale study constructed a text-based machine learning model to estimate replication probabilities across more than 14,000 psychology articles \cite{youyou2023discipline}. This work demonstrates that replicability varies meaningfully across subfields and methodological approaches, and that textual signals can help quantify this heterogeneity at scale. Such findings suggest that replication prediction can move beyond structured metadata toward richer representations derived from scientific narratives themselves.

\subsection{Artificial prediction markets}
Prediction markets are platforms to aggregate and disperse information into efficient forecasts of uncertain future events \cite{Arrow:2008, hanson2006information, manski2006interpreting, wolfers2006interpreting}. Participants in these markets buy and sell contracts representing outcomes of well-defined future events. The price of an asset in these markets can be understood as participants' collective prediction of the given outcome. Prediction markets have been successfully used, e.g., for sports betting \cite{spann2009sports}, election forecasting \cite{berg2008prediction}, infectious disease prediction \cite{polgreen2007use}, and aggregating employee wisdom in corporate settings \cite{cowgill2009using,gillen2012information}. A comprehensive review of prediction markets and their applications is provided in \cite{horn2014prediction}. 

Artificial prediction markets are numerically simulated markets populated by agent traders for the purpose of probability estimation. Theoretical work has offered support for the mathematical connections between artificial markets and machine learning algorithms \cite{chen2008complexity, storkey2011machine, abernethy2011optimization, hu2014multi}.  In initial formulations by Barbu and Lay \cite{lay2010supervised, barbu2012introduction,lay2012artificial,barbu2013artificial}, each agent is represented as a budget and a simple betting function. During training, each agent's budget is updated based on the accuracy of its predictions over a training dataset. An alternate market was proposed by Storkey and colleagues based on agents with isoelastic utilities \cite{storkey2012isoelastic}. Jahedpari et al. \cite{jahedpari2017online} introduced an online artificial prediction market that works like a regressor for adaptive trading strategies. They established that a continuous artificial prediction market could be useful for syndromic surveillance. Most recently, Nakshatri et al. \cite{nakshatri2021design} proposed an artificial prediction market wherein agent purchase logic is defined geometrically, in particular, by a convex semi-algebraic set in feature space. Time-varying asset prices affect the structure of the semi-algebraic sets, leading to time-varying agent purchase rules. Agent parameters are trained using an evolutionary algorithm. Authors show that their approach has desirable properties, e.g., the market satisfies certain universal approximation properties, and there exist sufficient conditions for convergence. Our work builds on this approach.

\subsection{Hybrid human-AI forecasting}
Researchers have increasingly emphasized the potential of hybrid intelligence, arguing that superior outcomes can emerge when human cognitive capabilities such as creativity, contextual reasoning, and ethical judgment are combined with the scalability and analytical power of artificial intelligence \cite{dellermann2019hybrid, wang2020human}. Prior work spans a broad range of efforts, from incorporating human factors into technology design \cite{bansal2019beyond, canonico2019wisdom, harper2019role} to leveraging human input for training data, as well as enabling post-hoc human interpretation and auditing in applications related to civic welfare \cite{fogliato2022case}, criminal justice \cite{travaini2022machine}, and beyond. The overarching aim of these efforts is to develop hybrid systems that effectively combine human intuition with machine rationality, enabling more efficient and robust approaches to complex societal challenges \cite{wang2019human, wang2020human, okamura2020adaptive}.

However, research on human-AI collaboration also highlights important limitations and design challenges. While AI generated inputs can enhance individual decision making performance, they may also unintentionally suppress unique human insights or reduce the diversity of perspectives that contribute to collective judgments. Fügener et al. \cite{fugener2021will}, for example, demonstrate that although AI advice can improve individual accuracy, it can simultaneously harm collective outcomes such as the wisdom of crowds. These findings illustrate that naive integration of AI advice into human decision processes may undermine the benefits of human diversity and expertise, underscoring the importance of carefully designed hybrid systems. %Within this broader context, human-AI forecasting research examines how human judgment and algorithmic forecasting models can be combined to produce predictions that are more accurate, scalable, and decision relevant than either component alone. Several studies have explored this collaboration in real-world forecasting tasks. For instance, 
A recent study proposes a human-AI collaboration framework for product demand forecasting, focusing on predicting future sales volume across different industries \cite{nair2024pair}. Their framework demonstrates that the optimal allocation of human and algorithmic input depends on factors such as product life cycle and demand volatility. Their empirical analysis shows that AI systems perform particularly well in high-frequency data environments, while human judgment remains essential for contextual interpretation and adjustments in rare or unexpected situations. Other studies highlight how human-AI interaction affects forecasting outcomes in practice. Fahse et al. \cite{fahse2023exploring} examine AI-assisted next day sales forecasting in a real-world bakery retail setting. Their study investigates how human decision makers interact with AI generated demand forecasts and how this interaction influences operational performance. Although AI support improved forecast accuracy and reduced product waste, it also led to lower overall sales. This finding illustrates that minimizing forecast error alone does not necessarily optimize real-world business outcomes, and that human oversight can be necessary to balance predictive accuracy with broader strategic considerations. Similarly, Revilla et al. \cite{revilla2023human} conduct a large-scale field experiment in retail demand forecasting to examine how different levels of human-AI collaboration influence predictive accuracy. Their findings show that the effectiveness of human intervention depends heavily on task uncertainty and the forecasting time horizon. In environments characterized by high uncertainty and short-term horizons, automated AI forecasting performs best. In contrast, human augmentation improves predictive accuracy in more stable environments with long term horizons. These results highlight the context dependent nature of human-AI complementarities in predictive tasks. Beyond commercial forecasting settings, hybrid forecasting approaches have also been applied to operational planning problems. Li et al. \cite{li2023integrated, li2024augmented} examine a scheduling approach for ship refit projects by integrating machine learning based predictions of task duration into a constraint programming optimization model. Their study compares expert estimates, historical data forecasting, and augmented human-AI forecasting approaches. The results show that combining human expertise with AI predictions produces more accurate task duration estimates and leads to more robust scheduling outcomes.%, reducing disruptions in resource allocation. Together, these findings suggest that hybrid forecasting systems can enhance decision making performance when human expertise and algorithmic predictions are effectively integrated.

\section{Data}
Algorithmic agents were trained on the outcomes of 402 replication studies of 
published findings in the social and behavioral sciences. In particular, the training dataset included outcomes of 313 studies assembled from the Reproducibility Project Psychology~\cite{nosek349corresponding}, the Social 
Science Replication Project~\cite{camerer2018evaluating}, the Experimental 
Economics Replication Project~\cite{camerer2016evaluating}, Many Labs~\cite{klein2014investigating}, 
and Many Labs 2~\cite{klein2018many}. The remaining 89 training datapoints leveraged replication studies run by the Center for Open Science as part of the Defense Advanced Research Project Agency (DARPA)'s Systematizing Confidence in Open Research and Evidence (SCORE) 
program,\footnote{See \url{https://www.darpa.mil/program/systematizing-confidence-in-open-research-and-evidence}.} 
spanning psychology, marketing, economics, sociology, 
political science, education, management, health, criminology, and public 
administration. Table \ref{domainstable} shows the breakdown of training data by field.

%\usepackage{caption}
%\captionsetup[table]{font=small}
%\captionsetup[table]{labelfont=bf, font=small}
\begin{wraptable}{r}{0.35\textwidth}
\centering
\captionsetup{labelfont={bf,footnotesize},textfont=footnotesize}
\caption{\footnotesize Replication outcomes used for training (total = 402), by domain.}
\footnotesize
\begin{tabular}{lc}
\hline
\textbf{Category} & \textbf{n} \\
\hline
Psychology & 252 \\
Economics & 99 \\
Marketing/Org Behavior & 20 \\
Sociology & 8 \\
Political Science & 6 \\
Education & 5 \\
Management & 5 \\
Health & 4 \\
Criminology & 2 \\
Public Administration & 1 \\
\hline
\end{tabular}
\label{domainstable}
\end{wraptable}

An additional 30 replication study outcomes from the SCORE project were held out as test data 
for the hybrid market experiments, with five studies evaluated per 
disciplinary session across six domains: economics, sociology, psychology, 
marketing/organizational behavior, political science, and 
education.\footnote{All train and test data is available at 
\url{https://osf.io/g5sny/overview}.} Although the SCORE replication outcomes are now publicly available, they were not yet published at the time of market experiments, so participants had no way to look them up. %The domains and sources of replication training and test data are detailed in Table1 and Table2.

\begin{figure}
    \vspace{-0.6cm}
    \centering
    \includegraphics[width=\textwidth]{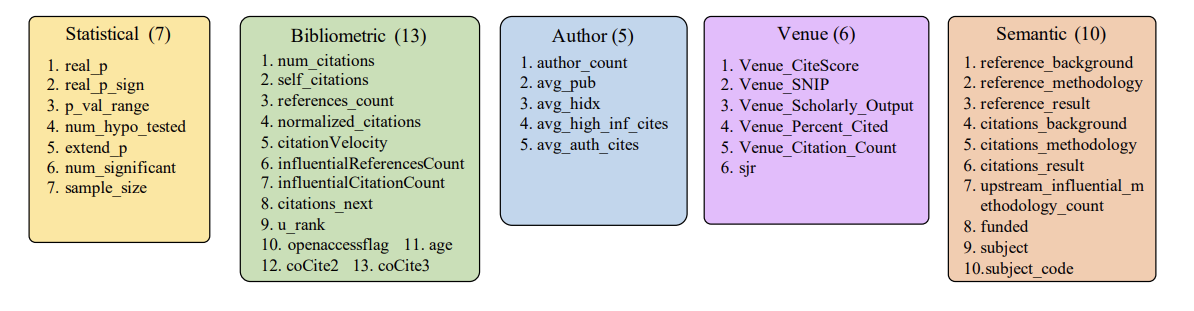}
    \caption{Forty-one features were extracted from full text and corresponding metadata for each research claim in the training and test datasets. (see \cite{wu2021predicting})}.
    \label{fig:five different features}
    %\vspace{-0.cm}
\end{figure}

The full text of both training and test papers, along with corresponding metadata, 
were processed through a feature extraction pipeline to derive semantic, 
bibliometric, and statistical features following the methodology described 
in~\cite{wu2021predicting}. In total, 41 features were extracted per research 
claim, including statistical evidence, author characteristics, publication 
venue metrics, funding acknowledgments, and related metadata 
(see Figure~\ref{fig:five different features}).

%\begin{figure}
%     \centering
%     \begin{subfigure}[b]{0.45\textwidth}
%         \centering
%         \includegraphics[width=\textwidth]{distribution_of_data_categories_for_training.png}
%         \caption{Category Totals}
%         \label{fig:category totals}
%     \end{subfigure}
%     \hfill
%     \begin{subfigure}[b]{0.45\textwidth}
%         \centering
%         \includegraphics[width=\textwidth]{distribution_of_data_categories_for_training (percent).png}
%         \caption{Category Totals (\%)}
%         \label{fig:category totals (percent)}
%     \end{subfigure}
%        \caption{Distribution of Training Publication Categories: 402 publications comprising 313 publications from Uzzi, Yang and 89 publications from TA3 datasets.}
%        \label{fig:training categories}
%\end{figure}

%\begin{table}[htbp]
%\centering
%\begin{minipage}{0.45\textwidth}
%\centering
%\caption{Replication Outcome PDFs (Total = 89)}
%\begin{tabular}{lc}
%\hline
%\textbf{Category} & \textbf{n} \\
%\hline
%Psychology & 30 \\
%Marketing / Organizational Behavior & 20 \\
%Economics & 8 \\
%Sociology & 8 \\
%Political Science & 6 \\
%Education & 5 \\
%Management & 5 \\
%Health & 4 \\
%Criminology & 2 \\
%Public Administration & 1 \\
%\hline
%\end{tabular}
%\end{minipage}
%\hfill
%\begin{minipage}{0.45\textwidth}
%\centering
%\caption{Uzzi \& Yang 2021 PNAS Dataset (Total = 313)}
%\begin{tabular}{lc}
%\hline
%\textbf{Subset} & \textbf{n} \\
%\hline
%RPP & 90 \\
%Econ Wiki & 88 \\
%Psychology\_Additional & 80 \\
%Psych FileDrawer & 52 \\
%Econ 18 & 3 \\
%\hline
%\end{tabular}
%\end{minipage}
%\end{table}

\section{Market Model}
Prior research has introduced artificial prediction markets composed of AI agents who buy and sell replication outcomes \cite{rajtmajer2022synthetic}. In these markets, model weights are optimized using genetic algorithms trained on ground-truth replication studies to evolve agent behaviors that predict reproducibility. Our hybrid markets adopt this foundational structure. 

\begin{wrapfigure}{r}{0.5\textwidth}
    \vspace{-0.4cm}
    \centering
    \includegraphics[width=0.5\textwidth]{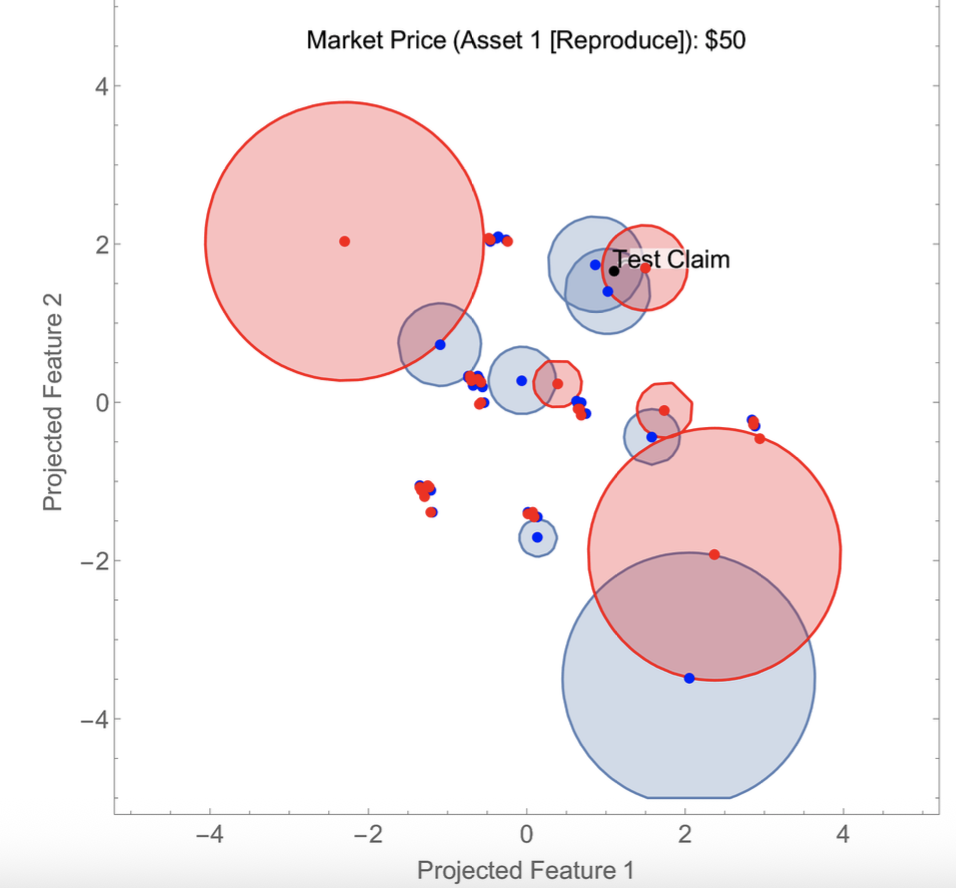}
    \captionsetup{labelfont={bf,footnotesize}, textfont=footnotesize}
    \caption{A toy artificial market with input data from the Reproducibility Project Psychology. Agents are represented as blue or red ellipsoids representing "will replicate" and "will not replicate" specialization, respectively. \emph{Note 1: High dimensional feature space is projected down for visualization. Note 2: We multiply the price by 100 and convert to dollars.}}    
    \label{fig:marketfig1}
    \vspace{-0.4cm}
\end{wrapfigure}

The mathematical construction of these markets is fully laid out in prior work \cite{nakshatri2021design}. At a high level, synthetic agents interact in a simple binary option market using a logarithmic market scoring rule. Agents begin with an initial amount of cash and may choose to purchase contracts representing "will replicate" or "will not replicate" outcomes of a given replication study (each market corresponds to one replication study). Agents are instantiated in feature space at the location of training data points and specialize in buying either "will replicate" or "will not replicate" contracts depending on the training datapoint they represent. Agents in the market bid in geometric regions of feature space (see Figure \ref{fig:marketfig1} for notional representation). Agents are sensitive to asset price, which causes their bid behavior to evolve in time.

%Each agent can be conceptualized as a multi-dimensional hyperparameter ellipsoid derived from the features of a publication in the training dataset. In this hyperparameter space, the radii of the ellipsoid correspond to the agent’s propensity to participate in the market, that is, its willingness to buy contracts associated with replicable or non-replicable outcomes. This behavior is governed by the model parameter percent difference, which defines how sensitive an agent is to deviations in feature space when deciding whether to trade. Once a trade occurs, the Liquidity parameter determines how rapidly an asset can be exchanged without significantly altering its price, thereby affecting overall market responsiveness. Lambda further modulates agent participation by controlling the frequency at which agents are activated in successive market increments; effectively, Lambda influences the size and dynamism of the active agent pool. In addition to these core parameters, the genetic algorithm includes two tunable settings, initial agent cash and market duration, which shape the broader trading environment and influence the evolutionary optimization process. 

In our hybrid markets, each agent was furnished with \$500 cash, a value that ensured that each agent would not exhaust its cash supply during the experiment. We chose this configuration to mimic real market forces, where there is consistent market momentum throughout the experiment. %Yet, with appropriate tuning of liquidity and lambda, the market can perceptibly move through aggregate human transactions. 
Human participants were each given \$25 to invest per market, with each market bookkeeping operating independently. The specific value balanced budgetary considerations with incentivization. Liquidity and lambda parameters were tuned with these amounts fixed, specifically, in order to support balance between agent and human influence. We imposed a minimum activity rule for human participants, namely, we required a minimum of three trades in order for a human participant to be eligible to receive their earned payout. %with the associated payout. The participants did not know in advance which paper was the money maker. 

Hybrid markets were configured to run for a total duration of 12 hours, corresponding to approximately 43,200 one-second increments. During live experiments, software latency reduced the effective number of increments to roughly 39,000. Agent weights were determined by the size of the training corpus and the number of agents instantiated in each market, such that the total weight reflected the product of the number of training papers and the number of agents participating per market.

\begin{figure}
    \vspace{-0.6cm}
    \centering
    \includegraphics[width=\textwidth]{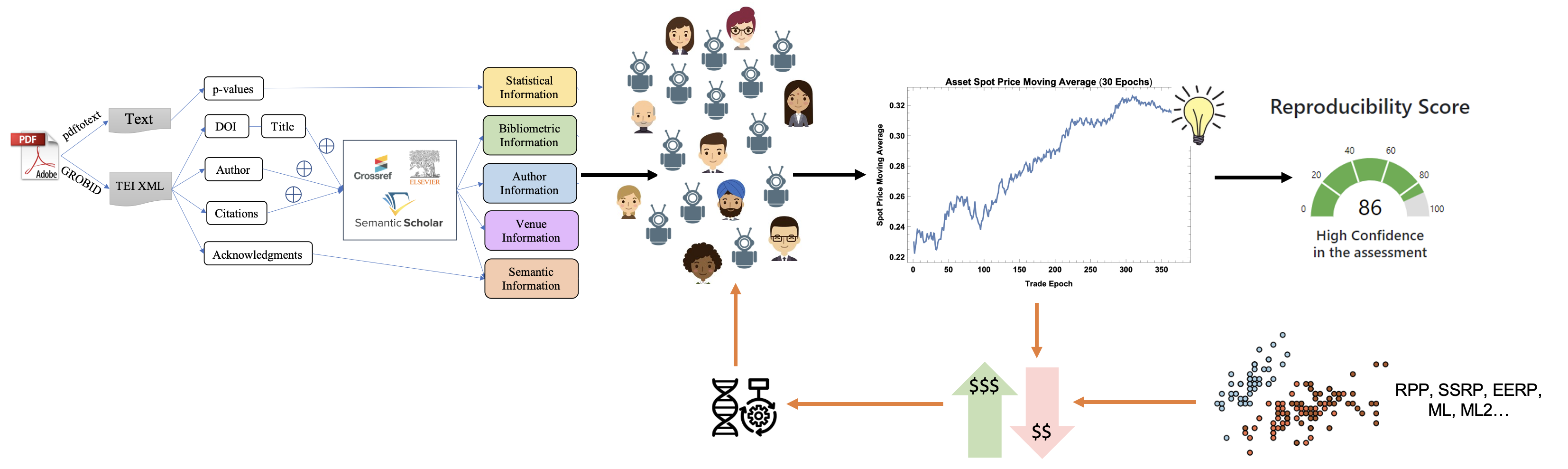}
    \caption{Schematic representation of hybrid markets for replication prediction. Step 1: Features are extracted from full text of a paper of interest. Step 2: Extracted features are passed to algorithmic traders; the full text PDF is provided to human traders. Step 3: Algorithmic and human traders buy and sell contracts representing "will replicate" or "will not replicate" outcomes of a replication study associated with the primary claim of the paper of interest. Step 4: Trading manipulates the underlying asset prices via logarithmic market scoring rule. Step 5. At market close, the price of a "will replicate" asset is interpreted as the market's confidence in the replicability of the given claim.  (\emph{training phase, orange arrows}) Algorithmic traders profit or lose money based on the total value of the assets they hold. Agents who profit are allowed to reproduce, mutate, and remain in the market via genetic algorithms.}
    \label{fig:schematic}
    %\vspace{-0.cm}
\end{figure}

\subsection{Market Training}

For hybrid experiments, we implemented a training methodology that first optimized the key hyperparameters of the system: lambda; liquidity; percent difference; initial agent cash, and market duration. The performance of each hyperparameter configuration was evaluated using two complementary criteria. The first metric assessed overall market accuracy, defined as the model’s ability to correctly predict ground-truth replication outcomes in the training dataset. Because multiple configurations may achieve similarly high predictive accuracy, a second evaluation criterion focused on the plausibility of agent participation patterns.

To assess whether a given hyperparameter configuration produced a realistic market, we examined resulting agent engagement behavior. Configurations that caused all agent ellipsoids to initialize with excessively small radii resulted in no agents identifying test data points as sufficiently close, leading to negligible trading activity. Conversely, configurations that produced excessively large radii caused all data points to appear similarly close to all agents, prompting universal participation and eliminating meaningful differentiation among agents. A viable and realistic configuration is therefore one that preserves meaningful inter-agent diversity across test data points and supports participation levels that reflect the categorical distribution and disciplinary variation present in the training corpus. This dual-metric approach allowed us to select hyperparameter settings that were both high-performing and behaviorally coherent.

%The following parameters were the final ones for the artificial prediction market: 
%\begin{itemize}
%    \item Percent: 0.0101 or 1.01\%
%    \item Liquidity: 60
%    \item Lambda: 0.4
%    \item Initial Agent Cash: \$15
%    \item Market Duration: 60
%\end{itemize}

\subsection{Market Configuration for Hybrid Experiments}
As noted, we adopted a one-second trading resolution for hybrid markets to approximate the temporal dynamics of traditional financial markets. This required adjustments to the artificial market’s training parameters in order to extend market convergence normally achieved within a short sequence of agent-only trades across a substantially longer timeframe. During artificial market training, the model was optimized using a market duration of 60 iterations. In contrast, the human-AI hybrid experiment was designed to run for 12 hours, corresponding to 43,200 one-second increments. To maintain continuity of agent participation under this extended schedule, we reduced the lambda parameter. Lowering lambda constrained the number of artificial agents eligible to trade at each increment, preventing premature convergence while preserving realistic market activity.

We also increased the cash allocation for agents to \$500 to ensure they remained solvent throughout the full 12-hour experiment. This modification helped stabilize artificial trading behavior and maintain consistent market dynamics across agents over the long duration of human participation. Together, these adjustments allowed us to better isolate the influence of human trading strategies by ensuring that AI agents continued to behave coherently and predictably across the extended duration of the hybrid market.

%The final parameters for the hybrid market are below: 
%\begin{itemize}
%    \item Percent: 0.0101 or 1.01\%
%    \item Liquidity: 60
%    \item Lambda: .001
%    \item Initial Agent Cash: \$500
%    \item Market Duration: 43200
%    \item Initial Human Cash: \$25
%    \item Initial Price: \$0.50
%\end{itemize}

\subsection{Hybrid Market Experimental Design}
\begin{figure}
    \vspace{-0.9cm}
    \centering
    \includegraphics[width=\textwidth]{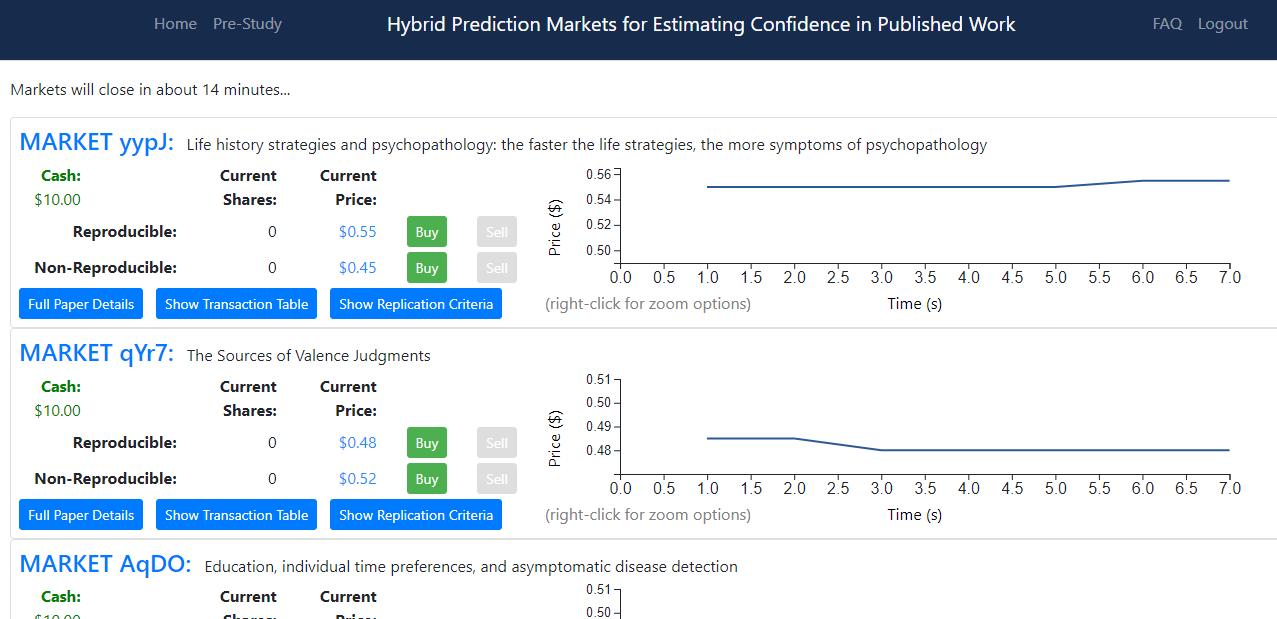}
    \caption{\textbf{Interactive web application.} Human participants are provided with information about the replication study and full text of the associated paper. We provide them with initial cash to invest. They may buy and sell contracts representing `will replicate' and `will not replicate' outcomes of the study via the app.}
    \label{fig:UI}
    \vspace{-0.3cm}
\end{figure}

In ongoing research, we have developed and beta-tested a platform that enables bot interactions with real human participants, i.e., a hybrid market environment, for replication prediction. The platform consists of a web server that hosts a pre-trained artificial market and provides several API endpoints that: (1) allow artificial agents to buy assets; (2) allow human participants to buy and sell assets; and (3) manage transaction bookkeeping and track experiment statistics. The system also includes an interactive web application that enables human participants to trade assets through an intuitive interface (see Figure \ref{fig:UI}). Although the artificial market is theoretically formulated as a continuous system, both the artificial-only and hybrid markets were implemented as discrete processes that updated every one second. Market transactions were handled using a queuing system that first processed all participating artificial agents, followed by any human transactions in a first-in, first-out order. Both artificial agents and human participants were restricted to single-share trades. After each transaction, asset prices were updated using the logarithmic market scoring rule \cite{nakshatri2021design}. 

Institutional Review Board (IRB) approval was obtained for all hybrid and human prediction markets prior to subject recruitment. Human participants were recruited between January and March 2023. In particular, we recruited researchers in economics, sociology, psychology, marketing, political science, and education. Individuals currently enrolled in a PhD program, University faculty, and PhD-holding individuals outside of academia were eligible to participate. We scraped the websites of universities in the United States to obtain email addresses and solicit researchers from relevant domains. In total, 97 researchers actively participated in this study. 

Hybrid market experiments were run as six separate events. Each of the six experiments involved five markets; that is, in each event, participants were given the opportunity to evaluate and buy or sell outcomes of five distinct replication studies. Participants were provided with the full text of the papers and the identified claims two days in advance of the market runs. All markets were held online and ran for 12 hours. For economics, sociology, and psychology claims, we had three types of experiments: human-only markets (no agents), hybrid human-AI markets, and artificial markets (agents alone). For marketing, education, and political science, we ran only hybrid and artificial markets. 

Each participant was compensated with \$40 for their participation and given \$125 to use in each of their five markets (\$25 in each). One of the five markets from each event was randomly selected as the "money market". Participants were allowed to keep their total cash from the money market at the end of the event. Participants were not informed which of the 5 was the money market until after markets closed.

\section{Experimental Results}

\subsection{RQ1: Hybrid-AI Forecasts}

Hybrid human-AI prediction markets yielded predictive performance comparable to or better than artificial markets across most domains, the exceptions being marketing and education. Hybrid markets achieved the lowest prediction error in sociology and political science and performed competitively in economics, while human-only markets performed best in psychology. These results suggest that combining algorithmic predictions with human expert judgment is a plausible approach to scientific forecasting.

\textbf{Economics}
Economics markets took place on 19th April 2023.  The human-only market outperformed both the hybrid and AI markets. The mean absolute error (MAE) was 0.414 for the human-only market, as shown in in Table \ref{tab:mae_by_discipline}. The hybrid market had a lower overall absolute error than the AI market, at 0.411 vs. 0.452. A total of 40 participants were involved in these markets; 20 out of 40 made a profit in the money market.

\textbf{Sociology}
Sociology markets were run on 24th April 2023. Both the AI and hybrid markets outperformed the human-only market. The hybrid market achieved substantially lower overall absolute error compared to both the human-only and AI markets, indicating the highest level of predictive accuracy in this domain (see Table \ref{tab:mae_by_discipline}). A total of 33 human participants took part in the event; 14 out of 33 made a profit in the money market.

\textbf{Psychology}
Psychology markets took place on 21st April 2023. The human-only market again performed better than both the hybrid and AI markets, recording 0.378 MAE (see Table \ref{tab:mae_by_discipline}). Absolute error was nearly identical for the hybrid and AI markets, indicating comparable predictive performance between these two approaches. 40 human participants took part in the event; only 12 made a profit in the money market. 

\textbf{Marketing}
Marketing markets were run on 12th April 2023. The artificial market outperformed the hybrid market slightly. The AI market correctly predicted three markets, whereas the hybrid market correctly predicted two (Table \ref{tab:Final_price_2markettypes}). A total of 33 human participants took part in this study; 11 were able to make a profit in the money market. %The experiment took place on 12th April 2023. The final prices calculated by all the markets are represented in Table3.

\textbf{Political Science}
Political Science markets took place on 14th April 2023. The hybrid markets demonstrated stronger performance than the artificial markets (see Tables \ref{tab:Final_price_2markettypes}, \ref{tab:mae_by_discipline}). %Absolute error was also lower in the hybrid market,suggesting higher overall prediction accuracy (see Table \ref{tab:mae_by_discipline}). 
The event included 37 human participants; 18 profited in the money market.

\textbf{Education}
Education markets took place on 17th April 2023. Both artificial and hybrid markets correctly predicted three outcomes of five. However, overall mean absolute error was slightly lower for the artificial market, indicating marginally better prediction (see Table \ref{tab:Final_price_2markettypes}). Experiments involved 31 participants, of whom 13 profited in the money market.  

\begin{table}[]
\centering
\begin{tabular}{|p{0.9in}|p{.4in}|p{.35in}|p{.35in}|p{.35in}|p{.4in}|p{.35in}|p{.35in}|}
\hline
Market & Final Price Artificial & Final Price Hybrid & Final Price Human & \textbf{Ground Truth} & Final Pred. Artificial & Final Pred. Hybrid & Final Pred. Human \\
\hline\hline
% -------------------- Econ --------------------
1574 (Econ) & 0.665 & 0.75  & 0.69 & \textbf{R}  & R & R & R \\ \hline
AgO1 (Econ) & 0.698 & 0.727 & 0.62 & \textbf{R}  & R & R & R \\ \hline
PIDa (Econ) & 0.708 & 0.737 & 0.48 & \textbf{R}  & R & R & NR \\ \hline
QIIV (Econ) & 0.67  & 0.683 & 0.43 & \textbf{NR} & R & R & NR \\ \hline
VB9K (Econ) & 0.661 & 0.586 & 0.43 & \textbf{NR} & R & R & NR \\
\hline\hline
% -------------------- Sociology --------------------
5Kgq (Sociology) & 0.63 & 0.78 & 0.76 & \textbf{NR} & R & R & R \\ \hline
dxQp (Sociology) & 0.67 & 0.77 & 0.77 & \textbf{R}  & R & R & R \\ \hline
e227 (Sociology) & 0.64 & 0.83 & 0.83 & \textbf{R}  & R & R & R \\ \hline
Vj0p (Sociology) & 0.66 & 0.71 & 0.75 & \textbf{NR} & R & R & R \\ \hline
x0pA (Sociology) & 0.62 & 0.77 & 0.23 & \textbf{R}  & R & R & R \\
\hline\hline
% -------------------- Psychology --------------------
521q (Psychology) & 0.648 & 0.741 & 0.61 & \textbf{NR} & R & R & R \\ \hline
88xa (Psychology) & 0.643 & 0.393 & 0.41 & \textbf{R}  & R & NR & NR \\ \hline
8wZ0 (Psychology) & 0.71  & 0.574 & 0.17 & \textbf{NR} & R & R & NR \\ \hline
Br0x (Psychology) & 0.714 & 0.903 & 0.94 & \textbf{R}  & R & R & R \\ \hline
EQxa (Psychology) & 0.637 & 0.594 & 0.46 & \textbf{NR} & R & R & NR \\ 
\hline
\end{tabular}
\footnotesize
\caption{Final value (in dollars) of a "will replicate" asset (contract) in artificial, hybrid, and human-only markets for economics, sociology, and psychology. \emph{R = "replicated"; NR = "not replicated"}}
\label{tab:Final_price_3markettypes}
\end{table}

\begin{table}[]
\centering
\begin{tabular}{|p{0.95in}|p{.4in}|p{.35in}|p{.35in}|p{.35in}|p{.4in}|p{.35in}|p{.35in}|}
\hline
Market & Final Price Artificial & Final Price Hybrid & Final Price Human & \textbf{Ground Truth} & Final Pred. Artificial & Final Pred. Hybrid & Final Pred. Human \\
\hline\hline
% -------------------- Marketing --------------------
5XEE (Marketing) & 0.61 & 0.35 & - & \textbf{NR} & R & NR & - \\ \hline
8w97 (Marketing) & 0.674 & 0.6   & - & \textbf{NR} & R & R  & - \\ \hline
EKBZ (Marketing) & 0.76  & 0.35  & - & \textbf{R}  & R & NR & - \\ \hline
G1Lr (Marketing) & 0.66  & 0.437 & - & \textbf{R}  & R & NR & - \\ \hline
N8pB (Marketing) & 0.712 & 0.711 & - & \textbf{R}  & R & R  & - \\
\hline\hline
% -------------------- Political Science --------------------
mxyQ (Political Sci) & 0.663 & 0.49  & - & \textbf{NR} & R & NR & - \\ \hline
qgWj (Political Sci) & 0.65  & 0.843 & - & \textbf{R}  & R & R  & - \\ \hline
wRvv (Political Sci) & 0.673 & 0.753 & - & \textbf{R}  & R & R  & - \\ \hline
xYbO (Political Sci) & 0.677 & 0.586 & - & \textbf{NR} & R & R  & - \\ \hline
z4dO (Political Sci) & 0.619 & 0.549 & - & \textbf{R}  & R & R  & - \\
\hline\hline
% -------------------- Education --------------------
BIRQ (Education) & 0.681 & 0.578 & - & \textbf{R}  & R & R  & - \\ \hline
Bixd (Education) & 0.69  & 0.642 & - & \textbf{R}  & R & R  & - \\ \hline
bY2A (Education) & 0.696 & 0.668 & - & \textbf{NR} & R & R  & - \\ \hline
Kybl (Education) & 0.622 & 0.683 & - & \textbf{NR} & R & R  & - \\ \hline
zqwm (Education) & 0.659 & 0.689 & - & \textbf{R}  & R & R  & - \\ 
\hline
\end{tabular}
\footnotesize
\caption{Final value (in dollars) of a "will replicate" asset (contract) in artificial and hybrid markets for marketing, political science, and education. \emph{R = "replicated"; NR = "not replicated"}}
\label{tab:Final_price_2markettypes}
\end{table}

\begin{table}[]
\centering
\footnotesize
%\captionsetup{labelfont={bf,footnotesize},textfont=footnotesize}
\begin{tabular}{|p{1.2in}|p{.9in}|p{.9in}|p{.9in}|}
\hline
Discipline & MAE Artificial & MAE Hybrid & MAE Human-only \\
\hline\hline

Econ        & 0.452 & 0.411 & 0.414 \\ \hline
Sociology   & 0.472 & 0.424 & 0.536 \\ \hline
Psychology  & 0.528 & 0.523 & 0.378 \\ \hline
Marketing   & 0.430 & 0.490 & - \\ \hline
Political SC & 0.480 & 0.386 & - \\ \hline
Education   & 0.458 & 0.488 & - \\

\hline
\end{tabular}
\caption{Mean Absolute Error (MAE) by discipline.}
\label{tab:mae_by_discipline}
\end{table}

\subsection{RQ2: Trading Strategies as Reported in Post-Experiment Surveys}
At the conclusion of market experiments, all participants were asked to complete a brief survey which asked about their experiences and decisions in the market. Most participants relied heavily on their prior beliefs about the replication likelihood of the claim in question. Many explicitly stated that they bought assets corresponding with claims they personally believed would replicate or that they perceived as “undervalued” by the market spot price. %This suggests that individual epistemic judgment, rather than purely market dynamics, was a dominant driver of behavior. 
A second common strategy involved trend-following and price observation. Some participants reported looking at past price movements or market trends before making purchases, indicating adaptive behavior in response to market signals rather than relying solely on initial beliefs. We also saw evidence of profit-oriented and tactical trading, with a few participants mentioning strategies such as selling immediately when in profit, balancing buying and selling, or timing trades based on availability during the day. However, these financially strategic behaviors were less frequently cited. %escribed compared to belief-based trading. 
Notably, several responses indicated limited strategic engagement. Some participants reported not having a clear strategy, being confused, or participating for the first time without a structured approach. %This heterogeneity in financial literacy and engagement may have influenced overall market efficiency. %Overall, the dominant and most interesting insight is that trading behavior was primarily belief-driven rather than purely speculative or arbitrage-based, with only moderate evidence of systematic profit-maximizing strategies.

\section{Conclusions and Limitations}
This works puts forward hybrid human-AI prediction markets as a plausible approach to scientific forecasting. Across multiple live experiments involving participants from diverse academic disciplines, hybrid markets consistently matched or outperformed artificial only and human only baselines in overall predictive accuracy. While performance varied by domain, the results suggest that combining algorithmic agents trained on large replication datasets with human contextual judgment can enhance reliability in complex evaluative tasks. Overall, this work provides empirical support for the value of thoughtfully designed human-AI collaboration systems in scientific decision-making and contributes insights for building hybrid intelligence frameworks that integrate computational efficiency with human expertise.

These initial studies faced several limitations. Hybrid market experiments were conducted on a relatively small set of replication studies in specified domains, which may limit their generalizability. %Although  algorithmic agents were trained on hundreds of prior replication outcomes, the live hybrid evaluations involved a smaller number of markets and participants per disciplinary event. 
Larger-scale deployments would be needed to assess robustness across additional fields and claim types. In addition, participant engagement varied across experiments. In several markets, a non-trivial number of participants made no transactions, suggesting heterogeneity in participation, strategic sophistication, or familiarity with prediction markets. This variability may have influenced market performance and the degree to which human judgment was effectively aggregated.

\bibliographystyle{abbrv}
\bibliography{main.bib}

\end{document}